\documentclass{webofc}

\usepackage[varg]{txfonts}   
\usepackage{hyperref}
\usepackage{url}
\usepackage{braket} 
\hypersetup{colorlinks=true,citecolor=blue,urlcolor=blue,linkcolor=blue}
\begin{document}
\title{TrackHHL: A Quantum Computing Algorithm for Track Reconstruction at the LHCb}
%
%

\author{\firstname{Xenofon} \lastname{Chiotopoulos}\inst{1,2}\fnsep\thanks{\email{xenofon.chiotopoulos@maastrichtuniversity.nl}} \and
        \firstname{Miriam} \lastname{Lucio Martinez}\inst{3}\fnsep\thanks{\email{Miriam.lucio@ific.uv.es}} \and
        \firstname{Davide} \lastname{Nicotra}\inst{2}\fnsep\thanks{\email{d.nicotra@maastrichtuniversity.nl}} \and
        \firstname{Jacco} \lastname{A. de Vries}\inst{2}\fnsep\thanks{\email{jacco.devries@maastrichtuniversity.nl}} \and
        \firstname{Kurt} \lastname{Driessens}\inst{1}\fnsep\thanks{\email{kurt.driessens@maastrichtuniversity.nl}} \and
        \firstname{Marcel} \lastname{Merk}\inst{2}\fnsep\thanks{\email{m.merk@maastrichtuniversity.nl}} \and
        \firstname{Mark} \lastname{H.M. Winands}\inst{1}\fnsep\thanks{\email{m.winands@maastrichtuniversity.nl}} 
}

\institute{Department of Advanced Computing Sciences, Maastricht University, Maastricht, The Netherlands
\and
           Gravitational Waves and Fundamental Physics, Maastricht University, Maastricht, The Netherlands
\and 
          Instituto de Fisica Corpuscular, Centro Mixto Universidad de 
          Valencia - CSIC, Valencia, Spain
          }

\abstract{In the future high-luminosity LHC era, high-energy physics experiments face unprecedented computational challenges for event reconstruction. Employing the LHCb vertex locator as a case study we investigate a novel approach for charged particle track reconstruction. The algorithm hinges on minimizing an Ising-like Hamiltonian using matrix inversion. Solving this matrix inversion classically achieves reconstruction efficiencies akin to current state-of-the-art algorithms. Exploiting the Harrow-Hassidim-Lloyd (HHL) quantum algorithm for linear systems holds the promise of an exponential speedup in the number of input hits over its classical counterpart, contingent on the conditions of efficient quantum phase estimation (QPE) and effectively reading out the algorithm's output. This contribution builds on previous work by Nicotra et al.~\cite{Nicotra_2023} and strives to fulfill these conditions and further streamlines the algorithm's circuit depth by a factor up to $10^4$. Our version of the HHL algorithm restricts the QPE precision to one bit, largely reducing circuit depth and addressing HHL's readout issue. Furthermore, this allows for the implementation of a post-processing algorithm that reconstructs event Primary Vertices (PVs). The findings presented here aim to further illuminate the potential of harnessing quantum computing for the future of particle track reconstruction in high-energy physics.}

\maketitle
\vspace{-0.6cm}
\section{Introduction}
\label{intro}

\begin{figure*}[h]
\centering
\vspace*{1cm}       
\includegraphics[width=10cm,clip]{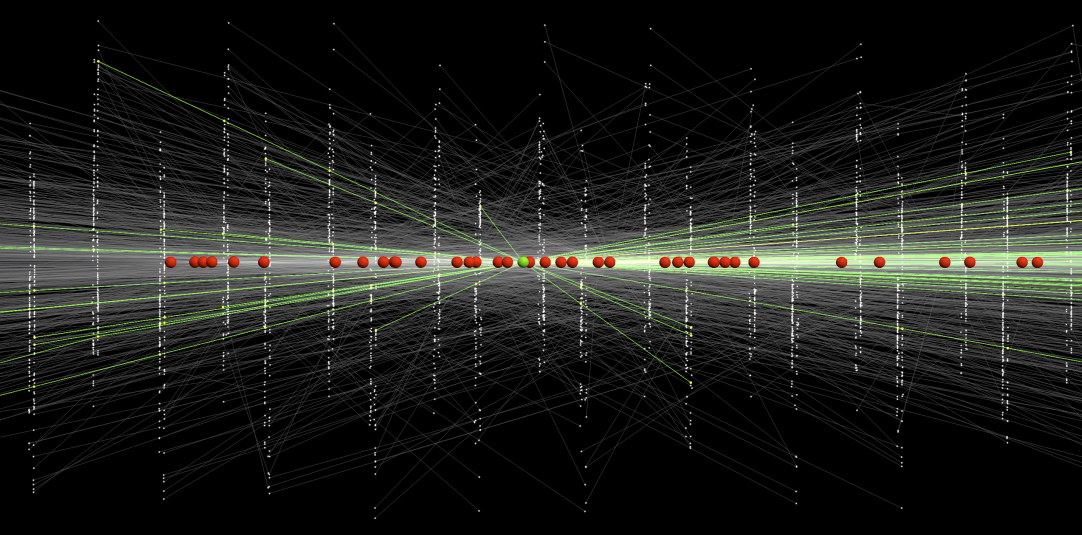}
\caption{Illustration of an LHCb event in the VELO detector, Created by Davide Nicotra}
\label{fig-2}       
\end{figure*}

 In the High Luminosity phase of the Large Hadron Collider (HL-LHC), thousands of particles are produced simultaneously and traverse sensitive detection layers where they deposit small amounts of energy, resulting in so-called \textit{hits} in the detectors. These hits are then reconstructed algorithmically into particle trajectories or \textit{tracks}, which are subsequently used in the determination of the collision points where the particles are produced: the Primary Vertices. A graphic display of an example collision event is shown in Fig.~\ref{fig-2}.


The increase of the number of simultaneous collisions in future colliders causes unprecedented volumes of highly complex data to be processed. Current software and computing hardware are expected to be insufficiently capable of dealing with such data volumes, because with increasing hit multiplicity a quadratic increase in combinatorics is required to reconstruct particle tracks. In an effort to explore novel solutions we investigate how quantum algorithms can be tailored to solve the track reconstruction problem in LHCb - in particular for the Vertex Locator (VELO). The VELO is the sub-detector closest to the LHCb collision point and, as it contains a negligible magnetic field, tracks can be modeled as straight line segments. As such, \textit{track} and \textit{primary vertex} reconstruction in the LHCb VELO detector provide a first case study.

There have been several attempts to leverage quantum computing for the track reconstruction problem in high-energy physics. Since tracking can be formulated as a Quadratic Unconstrained Binary Optimization problem, a wide range of quantum algorithms can be applied. Among these are Quantum Graph Neural Networks \cite{Tüysüz2021QGNN}, quantum annealing \cite{Zlokapa_2021DP_hamiltonian}, and the Variational Quantum Eigensolver \cite{tang2021qubit}. In addition, the tracking problem can be mapped to a linear system of equations. As shown by Nicotra et al.~\cite{Nicotra_2023}, this mapping satisfies all conditions required by the Harrow–Hassidim–Lloyd (HHL) algorithm \cite{Harrow_2009}, which theoretically provides an exponential speedup over classical methods in terms of computational complexity.

\vspace{-0.6cm}
\vspace{-0.4cm}

Building on this foundation, this work extends the HHL algorithm by introducing a specialized 1-Bit HHL variant. This variant reduces the number of qubits, circuit depth, and the number of measurement queries required to read out the final state, making the approach more feasible for future utility-scale quantum hardware. Although 1-Bit HHL reduces circuit depth and qubit requirements, a full track reconstruction still requires exhaustive sampling of the distribution to estimate it completely. To circumvent this readout problem altogether, we focus on extracting information on the primary vertices (PVs) instead. Once the HHL algorithm identifies active and inactive segments, we extrapolate the active segments to the LHC beam line and fit the resulting projections to find the PVs. This approach further reduces sampling from the final state as only a subset of segments is needed to reconstruct PVs.

\section{The LHCb Vertex Locator}
\label{VELO}

The Vertex Locator (VELO) is the first sub-detector in the LHCb spectrometer, positioned just $5 \ mm$ from the LHC beam line, and hence also from the proton–proton interaction point. It consists of 52 silicon pixel detectors installed in two detector halves, each side containing 26 modules. When the proton beams collide, they produce showers of charged particles traversing the silicon detector layers. These deposit energy in sensitive pixels, allowing the detector to record precise $(x,y)$ coordinates for each \textit{hit}. It is the goal of the reconstruction algorithm to group the hits into particle \textit{tracks} followed by track parameter estimation with a track fit. Subsequently, these tracks are used to identify the PVs. 


\section{Tracking with Matrix Inversion}
\label{Inversion}

In this paper, we build upon the tracking approach described by Nicotra et al.~\cite{Nicotra_2023}, based on a modified Denby-Peterson (DP) Hamiltonian:
\begin{equation}
    \mathcal{H}(\mathbf{S}) = \mathcal{H}_{ang}(\mathbf{S},\epsilon) + \alpha \mathcal{H}_{spec}(\mathbf{S}) + \beta \mathcal{H}_{gap}(\mathbf{S},N_{hits}) 
    \label{DP-hamiltonian}
\end{equation}
The track-finding problem is solved by building a system of linear equations where matrix inversion yields the solution of reconstructed tracks. 


In Equation \ref{DP-hamiltonian} the Hamiltonian is parametrized in terms of doublets $\mathbf{S}$, these doublets are possible connections between two hits in subsequent detector layers and take a binary value to indicate if they actively contribute to a track, $S_i \in \{0, 1\}$. Our implementation of the DP-Hamiltonian is explained in \cite{Nicotra_2023} and includes three terms. The angular term $\mathcal{H}_{ang}$ is the most important as it determines if a set of doublets $S_i$ and $S_j$ are considered to be aligned within $\epsilon$ and contribute to the Hamiltonian. 
\begin{equation}
    \mathcal{H}_{ang} (\mathbf{S}, \epsilon) = -\frac{1}{2} \sum_{a,b,c} f(\theta_{abc}, \epsilon) S_{ab} S_{bc}, \qquad
    f(\theta_{abc}, \epsilon) = \begin{cases}  1 & \text{if } \cos \theta_{abc} \geq 1 - \epsilon, \\ 0 & \text{otherwise}. \end{cases} 
    \label{angular_term}
\end{equation}

\noindent The remaining terms are given by $\mathcal{H}_{spec}$, which is quadratic in $S_i$, and 
$\mathcal{H}_{gap}$, which includes both quadratic and linear contributions in $S_i$:

\begin{equation}
    \mathcal{H}_{spec} (\mathbf{S}) = \alpha \sum_{ab} S^2_{ab}, \quad \quad \mathcal{H}_{gap} (\mathbf{S}) = \beta \sum_{ab} (1 - S_{ab})^2
    \label{spec_gap}
\end{equation}

 \noindent These terms are included for regularization, ensuring the resulting system of linear equations is positive semidefinite, so that its inversion does not lead to unexpected behavior. Nicotra et al.~\cite{Nicotra_2023} showed that by relaxing $S_i \in \mathbb{R}$, we find its minimum by taking the derivative of the quadratic $\mathcal{H}$ obtaining a system of linear equations: 

\begin{equation}
    \nabla \mathcal{H} = -A \mathbf{S} + b = 0
\end{equation}
The matrix $A = A_{ang} + A_{spec} + A_{gap}$ obtained by differentiating the quadratic terms and $b$ is defined by the linear terms $\mathcal{H}_{gap}$, chosen such that $b=\beta(1,1,...,1)$. The resulting vector $\mathbf{S}$ of real values is subsequently discretized to obtain an "on"/"off" status by setting a threshold $T$. The simplest example of this system is illustrated in Fig.~\ref{tripple}. For a more realistic case, this tracking method was classically evaluated on LHCb Run-3 simulated $B_s \rightarrow \phi \phi$ data and the achieved results are comparable with the current state of the art, see Ref~\cite{Nicotra_2023}.

\section{Quantum Track Reconstruction}
\label{Quantum-Tracking}

Given that the track reconstruction task can be approached by a matrix inversion,  it serves as an interesting use case for the HHL algorithm \cite{Harrow_2009}, which promises complexity improvements over the best classical alternatives. The matrix $A$ is Hermitian, invertible, and sparse, with a maximum condition number of $k=5$, due to the step function introduced in the Hamiltonian construction limiting the eigenvalue spread explored in Ref~\cite{Nicotra_2023}. The run-time complexity scales as $O(\kappa^2 \log N)$, compared to the best classical alternative of $O(\sqrt{\kappa} N)$ \cite{Harrow_2009}.

\subsection{Toy Model}
\label{Toy}

\begin{figure*}[h]
\centering
\vspace*{1cm}       
\includegraphics[width=13cm,clip]{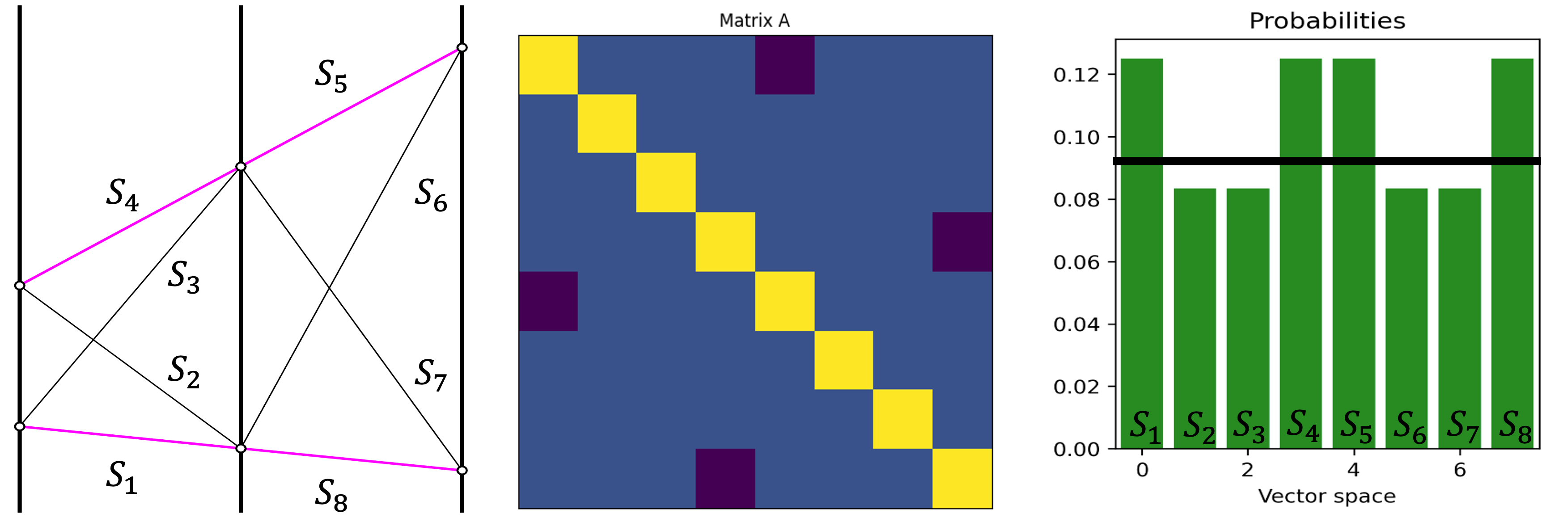}
\caption{The $\mathbf{Left}$ panel represents a graph construction of the minimal event, with each doublet labeled. Pink lines indicate active segments. The $\mathbf{Middle}$ panel is a heat-map of the matrix $A$ for the same event as the left panel. The $\mathbf{Right}$ panel illustrates the output distribution vector $S$ after inverting the matrix, where threshold T is applied to distinguish active from non-active segments.}
\label{tripple}       
\end{figure*}

As a testbed for developing this algorithm we used a simplified event model to explore the algorithm within the constraints of current quantum- hardware and simulations \cite{Nicotra_TrackHHL}. This toy model allows us to define custom VELO geometries, where the user can define the number of detector layers, particles, PVs and Hamiltonian parameters such as $\alpha$, $\beta$ and $T$. This Toy model assumes ideal detector conditions i.e. $100\%$ detector efficiencies, perfectly straight tracks and no scattering of particles due to passage through material. By adjusting parameters we can tune the toy and test for inefficiencies, limited resolution effects and kinked trajectories due to multiple scattering. For example, Fig.~\ref{tripple} showcases the simplest toy model event with 2 particles, 3 detector modules. The matrix is defined from the graph structure and the resulting $\mathbf{S}$ vector is then discretized by applying an appropriate threshold, see Fig~\ref{tripple}.

\vspace{-0.8cm}
\vspace{-0.8cm}

\subsection{Tracking with HHL}
\label{HLL}

The HHL algorithm \cite{Harrow_2009} consists of the following five steps:



\begin{enumerate}
    \item \textbf{State Preparation}: Represent the vector $\vec{b}$ as a quantum state $\ket{b}$. This assumes $\vec{b}$ can be efficiently encoded into a quantum register, which is trivial in our case and can be encoded with Hadamard gates.
    
    \item \textbf{Quantum Phase Estimation (QPE)}: Perform Quantum Phase Estimation on the matrix  $A$ . QPE estimates the eigenvalues  $\lambda_i$  of  $A$  and creates an entangled state: 
    \begin{equation}
        \sum_i c_i \ket{\lambda_i} \ket{u_i}
        \label{Step2}
    \end{equation}
    Where  $\ket{u_i}$  are the eigenvectors of  $A$ , and  $c_i$  are coefficients determined by  $\vec{b}$.
    
    \item \textbf{Eigenvalue Inversion}: Apply a controlled rotation to an ancillary qubit based on the inverse of the eigenvalues  $\lambda_i$. This transforms the state as follows:
    \begin{equation}
        \sum_i c_i \frac{1}{\lambda_i} \ket{\lambda_i} \ket{u_i}
        \label{Step3}
    \end{equation}

    \item \textbf{Uncomputation of QPE}: Reverse the QPE process to disentangle the eigenvalue register, leaving the state:
    \begin{equation}
        \sum_i c_i \frac{1}{\lambda_i} \ket{u_i}
        \label{Step4}
    \end{equation}

    \item \textbf{Measurement}: Measure the quantum state to extract information about the solution $\vec{S}$, which is encoded in the amplitudes of the quantum state. The ancilla qubit’s measurement is post-selected to ensure the controlled rotation was successful
\end{enumerate}

\subsection{Optimizing HHL}
\label{Optimizing}

The HHL algorithm implementation confronts us with two main challenges, the first being that QPE results in prohibitively deep circuits making it unfeasible for any hardware short of fault tolerant quantum computers. The second issue is that extracting the full spectrum of the final state information, as seen in Fig. \ref{tripple}, would need a prohibitive number of queries to the system to reconstruct this distribution, since {\bf all} track segment combinations in the vector space must queried. 


\begin{table}[ht]
\centering
\begin{tabular}{|r|r|r|r|r|r|r|}
\hline
\textbf{Optimization Type} & \textbf{Layers} & \textbf{Particles}  & \textbf{Qubits} & \textbf{Total gates} & \ \textbf{2-qubit gates}  \\ \hline
None                   &3                & 5         & 14                            & 14 515 229     & 7 107 317            \\\hline
Qiskit O3              &3                & 5         & 14                            & 1 089 295      & 501 066              \\\hline
Suzuki Trotter         &3                & 5         & 14                            & 257 367        & 120 171              \\\hline
1-Bit Phase Estimation &3                & 5         & 14                            & 11 547         & 8 780                \\\hline

\end{tabular}
\caption{Comparison of optimization methods for HHL shown here for a simulation of 5 particles traversing 3 layers, transpiled for the \texttt{ibm\_hanoi} device. Note here the 1-Bit Phase Estimation also incorporates the Suzuki Trotter optimization, and we will be referring to this as the 1-Bit HHL algorithm.}
\label{tbl:qiskit-results}
\end{table}
\vspace{-0.5cm}

A first pragmatic step at reducing the circuit size was to use optimization methods that were included as part of Qiskit \cite{qiskit2024}; in particular to use the optimization level 3 (O3) flag in the transpilation process. To benchmark the circuit depth and performance of our HHL implementation we transpiled to the $ibm\_hanoi$ $27-qubit$ $r5.11$ Falcon quantum processor. The use of O3 optimization results can be seen in Table \ref{tbl:qiskit-results} and achieved some degree of improvement in circuit depth without loss of performance, although still far from what is feasible in hardware.

\begin{figure*}
\centering
\vspace*{1cm}       
\includegraphics[width=13cm,clip]{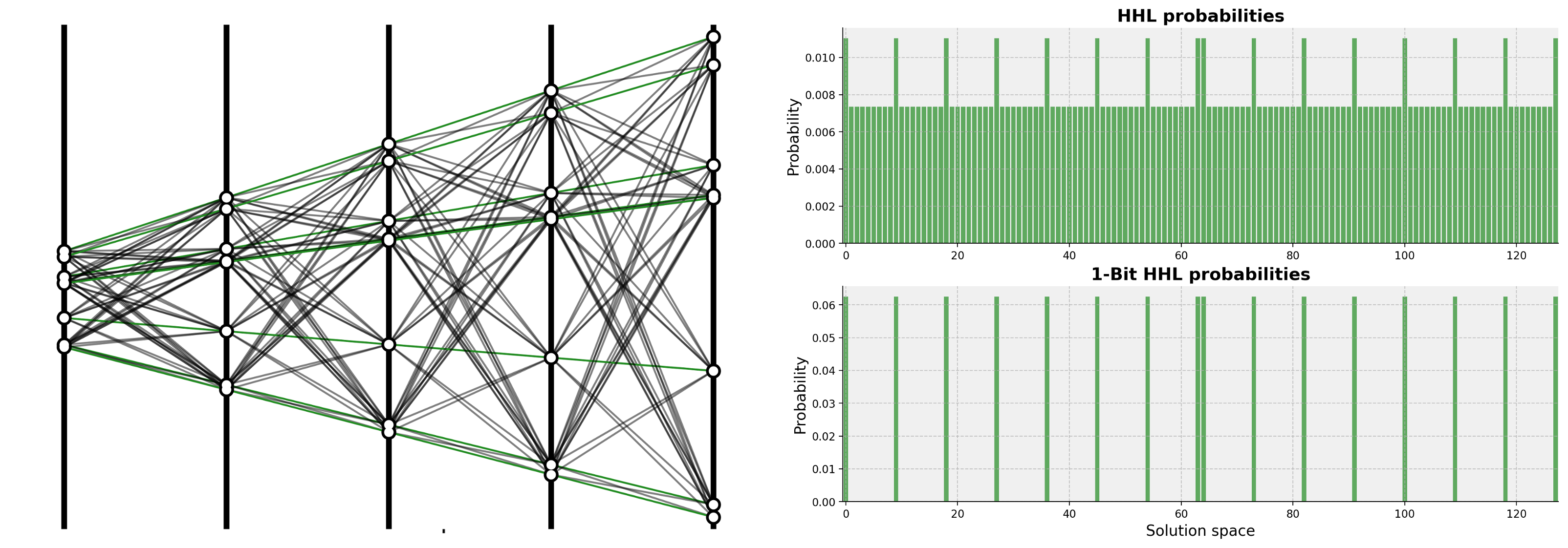}
\caption{On the $\bold{left}$  is an event with 5 layers and 8 particles, our largest simulated event with the corresponding distribution for 1-Bit HHL on the $\bold{right}$. Where all of the inactive segments are suppressed. }
\label{dist}       
\vspace{-0.4cm}
\end{figure*}

The second optimization attempt aimed to address the readout problem by leveraging the binary nature of the desired information, representing the probability whether a given doublet actively contributes to a track. Consequently, the use of the standard HHL algorithm, which outputs a full state vector with varying amplitude levels, is unnecessarily complex for this purpose. Instead, we applied a first-order Suzuki-Trotter Decomposition \cite{Ostmeyer_2023} to approximate the time evolution operator. The Suzuki-Trotter decomposition approximates the exponential as a sum of non-commuting operators $A$ and $B$ Fig. \ref{Suzuki-Trotter}, given $A$ and $B$ are distinct decomposed parts of the total Hamiltonian which are easier to exponentiate individually. Where $n$ represents the number of Suzuki-Trotter steps, and $t$ represents the simulated evolution time.

\begin{equation}
    e^{t(A + B)} \approx \left( e^{tA / n} e^{tB / n} \right)^n.
    \label{Suzuki-Trotter}
\end{equation}

The Suzuki-Trotter decomposition leads to shorter circuit depths, and using the coarse values $n=1$ and $t=1$, causes a shift in the final probability spectrum resulting in a suppression in non-active segments as seen when comparing the HHL and 1-Bit HHL distributions in Fig. \ref{dist}. This allows us to fully discriminate between active and non-active segments, effectively suppressing all contributions from non-active segments.

The final optimization leverages the binary nature of the problem to simplify the QPE process. HHL offers an advantage over classical algorithms by employing QPE to estimate the phases of the eigenvalues of the embedded Hamiltonian. However, given that the system under consideration is effectively two-level, and the lower eigenvalue is nearly zero due to the Suzuki-Trotter implementation, only a single bit of precision is required. By restricting the QPE accuracy to 1-Bit, we can efficiently determine whether a phase is close to zero or significantly different. This modification enables us to identify whether a segment is active.

\subsection{Implications of Optimization}
\label{Implications}

One of the most significant outcomes of the applied optimization methods is a substantial reduction in circuit depth by a factor of $10^3-10^4$ depending on the event. This brings us much closer to circuit depths feasible on current hardware. 

\vspace{-0.8cm}
\begin{figure*}[h]
\centering
\vspace*{1cm}       
\includegraphics[width=13cm,clip]{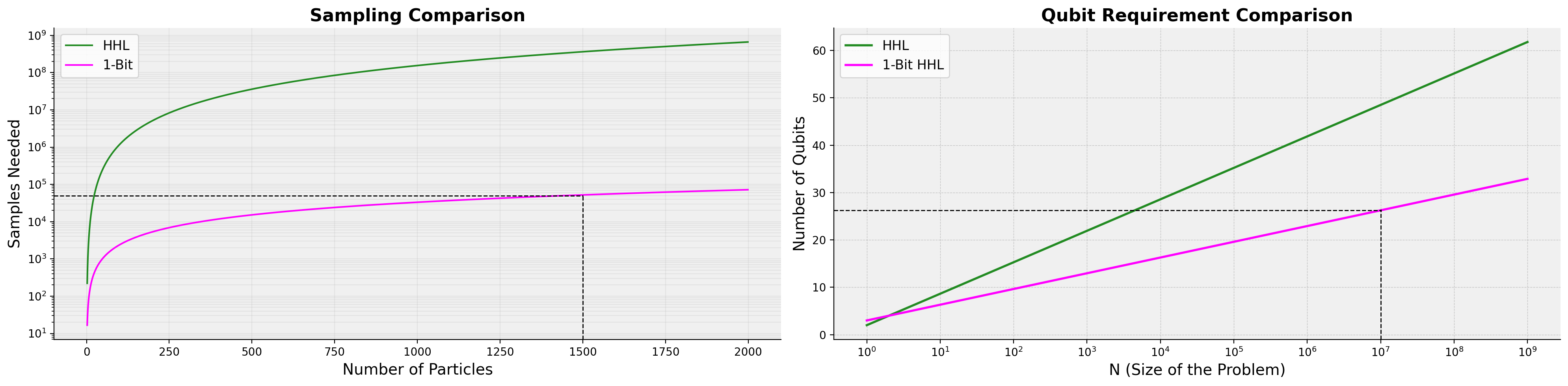}
\caption{The $\mathbf{Left}$ panel shows a graph of the number of samples needed to reconstruct the vector $S$ as a function of particles and 1-Bit HHL versus HHL. With dotted lines drawn at the HL-LHC requirements. The $\mathbf{Right}$ panel shows how the number of qubits needed when comparing 1-Bit HHL and HHL, with a similar dashed line drawn as the graph on the left, where the problem size $N$ relates to the number of particles as $N=N_p^2 N_{hits}$, given the number of particles $N_p$ and hits per particle $N_{hits}$.}
\label{output}       
\end{figure*}
\vspace{-0.4cm}

Another outcome is the alteration of the output distribution; as the non-active segments are suppressed we see an exponential reduction in the number of queries needed to the quantum system. The reduction of queries needed to the quantum system can be seen in Fig.~\ref{output}, and for conditions expected in the HL-LHC, i.e. 1500 particles per event, we see that instead of tens of millions now tens of thousands queries are needed to sample the full final state. This is due to the fact that the total number of segments scales with  $\mathcal{O} (N_p^2 N_{hits})$ for $N_p$ particles and average number of hits per particle $N_{hits}$, whereas the number of active segments only scales as $\mathcal{O}(N_p N_{hits})$. This leads to the scaling of the number of samples needed to reproduce the output distribution being $\mathcal{O}(N_p^2 N_{Hits} log (N_p^2 N_{Hits}))$. Whereas when we suppress the non-active segments the number of samples needed only scales with $\mathcal{O}(N_p N_{Hits} log (N_p N_{Hits}))$. This represents a significant enhancement, as under HL-LHC conditions less than $10^{5}$ queries to the quantum computer are required to capture all segments, as shown in Fig.~\ref{output}. A further reduction can be obtained with post-processing as discussed in Section \ref{Post-Processing}.

The second consequence of the single bit phase estimation is a reduction in the number of qubits needed. Since only one bit of precision is required to determine whether a segment is active, reduces the qubit scaling from $\mathcal{O}(2 log_2 N + 2)$ to $\mathcal{O}(log_2 N + 3)$ where $N$ is the size of the matrix A. This reduces our qubit requirement from $47$ qubits to $26$ qubits for a typical HL-LHC event, as seen in Fig.~\ref{output} and would be much more realistic in near-future hardware. 

\vspace{-0.3cm}
\section{Post Processing}
\label{Post-Processing}
\vspace{-0.15cm}

An approach to further mitigate the readout problem, is to combine segment information into direct reconstruction of Primary Vertices. To fully reconstruct the particle tracks we need to query the quantum system until we have a complete active segment sample and subsequently combine them into tracks using their geometric coordinates. 

\vspace*{-0.8cm}
\begin{figure*}[h]
\centering
\vspace*{1cm}       
\includegraphics[width=13cm,clip]{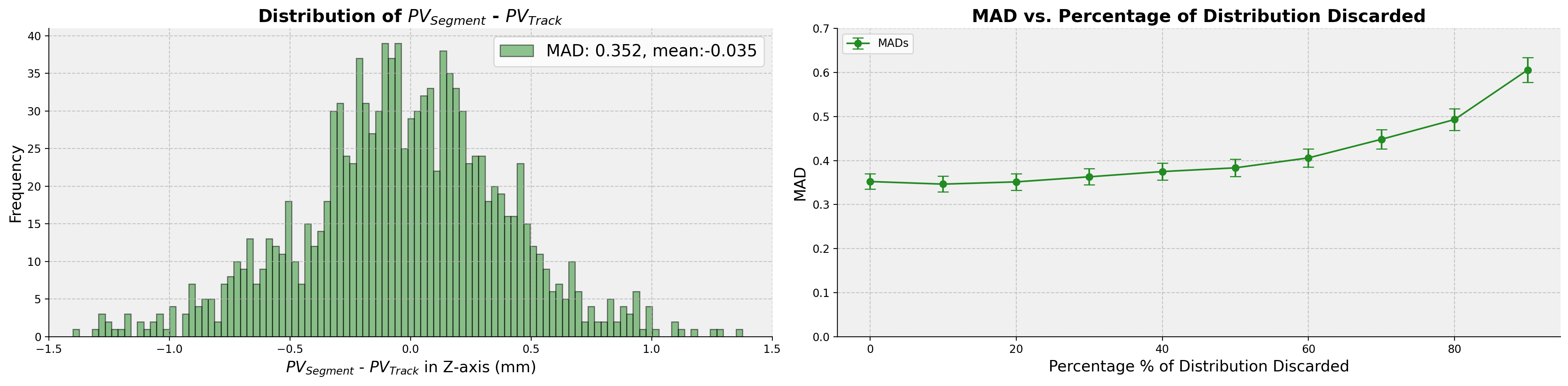}
\caption{The graph on the $\bold{left}$ shows the distribution of the difference between $PV_{Tracks}$ and $PV_{Segment}$. The Graph on the $\bold{right}$ shows how the Mean Absolute Difference changes as a function of the percentage of the distribution discarded. Generated using the same $B_s \rightarrow \phi \phi$ data as \cite{Nicotra_2023} }
\label{postprocessing}       
\end{figure*}
\vspace{-0.3cm}

Instead we attempt to reconstruct the PV with only a subset of the total segments. In the LHCb PVs are reconstructed by extrapolating VELO tracks to the beam-line at the coordinates $x=y=0$ followed by a histograms fitting procedure in $z$ \cite{Kucharczyk:1756296_PV}. Instead here, we extrapolate each active segment to the beam-line. To test the size of the subset of the distribution needed to reconstruct PVs we apply a DBSCAN \cite{DBSCAN} clustering to identify histogram peaks for both the segments and track extrapolation, and quantify the difference between the $PV_{Track}$ and $PV_{Segment}$. The resulting distribution is shown on the left of Fig.~\ref{postprocessing}, from which we calculate the Mean Absolute Difference (MAD). We repeat this procedure, while each time removing a larger fraction of the segments randomly, discarding an incremental amount of segments. The resulting MAD is plotted vs. the fraction of discarded track segments  on the right graph in Fig.~\ref{postprocessing}. From these figures we can see that discarding up to $60\%$ of our distribution only leads to minimal change in MAD value, allowing for a reduction in samples needed further improving the quantum state readout.  

\vspace{-0.3cm}

\section{Conclusion and Future Research}
\label{Conclusion}

In this work, we have demonstrated that track reconstruction in the LHCb's Vertex Locator can be reformulated as a system of linear equations, making it amenable to quantum algorithms such as HHL. Our results indicate that the naive application of HHL, while promising in theoretical complexity, faces practical limitations due to large circuit depths, extensive qubit requirements, and the high cost of fully reading out the quantum state. To address these, we propose a 1-Bit variant of HHL that drastically reduces the number of required qubits and circuit depth by exploiting the underlying binary nature of the reconstruction problem. Additionally, we introduce a Suzuki-Trotter decomposition to approximate the matrix exponential, suppressing all non-active segments, significantly reducing the amount of queries needed to reconstruct the final state distribution. Using this we then propose a post-processing method to further reduce the number of queries by reconstructing PVs rather than running full track reconstruction. For future research, we aim to test our minimal toy models on hardware and probe whether using the Quantum Singular Value Transform \cite{Gily_n_2019} would be appropriate for our use case.
\vspace{-0.2cm}
\subsubsection*{Acknowledgments} 
\label{Acknowledgments}
This publication is part of the project Fast sensors and Algorithms for Space-time Tracking and Event Reconstruction (FASTER) with project number OCENW.XL21.XL21.076 of the research programme ENW - XL which is (partly) financed by the Dutch Research Council (NWO).

\vspace{-0.01cm}

%
\bibliography{bib.bib}

\end{document}